\documentclass[aps,pra,showpacs,showkeys,twocolumn,longbibliography]{revtex4-1}

\usepackage{graphicx}
\usepackage{latexsym}
\usepackage{amsmath}
\usepackage{amssymb}
\usepackage{amsfonts}
\usepackage{dsfont}
\usepackage{color}
\usepackage[dvipsnames]{xcolor}

\begin{document}
\title{Taming the entanglement in the dynamical theory of weakly interacting Bose gases}

\author{Michiel Wouters }
\affiliation{TQC, Universiteit Antwerpen, Universiteitsplein 1,
B-2610 Antwerpen, Belgium}

\begin{abstract}

I show that the dynamics of the weakly interacting bose gas can be described by a modified time dependent Bogoliubov theory. The novelty of the approach is to include decoherence steps that gradually transform the entanglement entropy of the pure state into the von Neumann entropy of a statistical mixture. This approximation drastically reduces the entanglement that is needed in order to represent the system's state while becoming exponentially accurate in the mean field limit.
I argue that this scheme can be extended to all quantum systems whose ground state can be well approximated by a variational wave function. The upshot is that the dynamics of almost all quantum systems can be reduced to stochastic classical motion supplemented with small quantum fluctuations.
\end{abstract}
\date{\today}

\maketitle

\section{Introduction}

It should perhaps be more discomforting than generally acknowledged that no well-controlled theoretical treatment exists for the general dynamics of the weakly interacting Bose gas.
The determination of its ground state properties by Bogoliubov \cite{bogoliubov1947theory} was among the first successes of the application of second quantization to the quantum many body problem and the weakly interacting Bose gas is arguably the simplest nontrivial quantum many body system. The lack of a general strategy to deal with its dynamics reflects the immaturity of our understanding of the dynamics of quantum systems as compared to their ground state or thermal equilibrium properties. The huge difference in tractability of dynamics versus equilibrium  -- unique to the quantum realm and absent in classical physics -- is also conspicuous in tensor network approaches \cite{schollwock2011density}. They represent ground states very well, but struggle with  dynamics because of the ``entanglement wall'' \cite{calabrese2005evolution,kim2013ballistic}.


As a first approximation to the dynamics of the weakly interacting Bose gas, the classical Gross-Pitaevskii equation (GPE) has been very successful for the description of experiments with dilute ultracold atoms \cite{pitaevskii2016bose}. Problems arise however when one wishes to go beyond this purely classical approximation, i.e. when one wants to incorporate the discreteness of the atoms.
Among the most popular techniques that go beyond the GPE is the truncated Wigner approximation (TWA) \cite{sinatra2002truncated,polkovnikov2010phase}. It incorporates quantum fluctuations by sampling the initial Wigner distribution, but, the subsequent time evolution being governed by the GPE, it suffers from the UV catastrophe of classical field theories. The TWA can therefore only be used for short times or for systems where all modes are macroscopically occupied (deep in the classical regime). For longer times and in the quantum regime, it is inadequate.
The Wigner function is only one phase space representation of the Bose field, but approximations based on other representations ($P$, $Q$, positive $P$) all have their own problems and the TWA seems to be the more robust phase space approach.

The fact that phase space methods provide a good description for the largely occupied modes, combined with the almost ideal quasiparticle nature of the weakly occupied modes has led to several approaches where the mode space is separated into two regions \cite{davis_wright_gasenzer_gardiner_proukakis_2017,gardiner2000quantum,bijlsma2000condensate,zaremba1999dynamics}. The low energy modes (condensate region) are then described within the classical field description where the high energy modes (thermal cloud) are treated as a reservoir. The coupling between the condensate and the thermal cloud then leads to dissipative dynamics for the condensate field. The treatment of these coupled systems is however a formidable challenge and to the best of my knowledge has not been addressed without further drastic assumptions. These methods, that have been quite successful in modeling cold atom experiments, therefore rather enjoy phenomenological than fundamental status. 

It is thus safe to say that no controlled approach exists for the description of the dynamics of the weakly interacting Bose gas that becomes asymptotically exact in the mean field limit (interaction strength tending to zero, particle density tending to infinity). This is in stark contrast to the ground state properties, that are excellently described by Bogoliubov theory, that is closely related to a variational approximation within the Gaussian state manifold \cite{guaita2019gaussian}. Gaussian states are characterized by the expectation values of the annihilation operators (think for concreteness about a Bose-Hubbard model) $\alpha_j = \langle \hat a_j\rangle$ and the correlation functions $ \langle \delta  \hat  a^\dag_i \delta \hat a_j \rangle$ and $\langle \delta \hat a_i \delta \hat a_j \rangle$, where $\delta \hat a_j = \hat a_j - \alpha_j$, where the former represent the classical field and the latter the quantum fluctuations.   

The reason why the time dependent Bogoliubov theory fails is that quite generically, the fluctuations grow quickly under time evolution, while the validity of the approximation is only ensured when the fluctuations are small. When starting from a state with small fluctuations, time dependent Bogoliubov theory is then accurate only for short times, and it typically breaks down even before the TWA. It is the aim of this work to modify the Gaussian approximation in order to extend its validity to long times and I will argue that this method becomes asymptotically exact in the mean field limit.

There is a direct connection between the growth of the quantum fluctuations in the Gaussian theory and dynamical instabilities in the classical GPE. Indeed, the Gaussian (Bogoliubov) theory essentially comes down to the replacement  $\hat a_j \rightarrow \alpha_j+\delta \hat a_j$, where the classical amplitudes $\alpha_j$ obey GPE dynamics with some small modifications and the quantum fluctuations $\delta \hat a_j$ are in first approximation governed by the linearized dynamics around the GPE solution \cite{castin1998low}. 
When the linearized dynamics is stable, the fluctuations remain small and time dependent Bogoliubov theory is accurate at all times.
On the other hand, for unstable linearized dynamics, the quantum fluctuations grow exponentially.
Because the Gaussian variational approach conserves the total particle number, the growth of the density of quantum fluctuations leads to a large depletion of the classical field. At long times, the Gaussian theory then treats a large part of the system as quantum fluctuations, but this essentially means that the nonlinearities in the system are, apart from some mean field shifts, neglected.

The physical reason for the disappearance of the field expectation values $\alpha_j$ is that for a chaotic system in the presence of quantum fluctuations, the system evolves to a superposition of states with very different classical field amplitudes, that largely average out. 
The coherence of this superposition is however expected to be irrelevant, because no physically accessible observables exist that can distinguish between the coherent superposition and the incoherent mixture. Therefore, intuition suggests that discarding the coherence should constitute a negligible error. Under this approximation, the state of the system can be described as a classical mixture of Gaussian states that all have small quantum fluctuations. 
This strategy was already adopted in previous works \cite{wouters2020quantum,fernandes2022gaussian}, and it is the aim of the present paper to refine it in order to construct a well controlled approximation to the dynamics of the weakly interacting Bose gas.

\begin{figure}
    \centering
    \includegraphics[width=0.9\columnwidth]{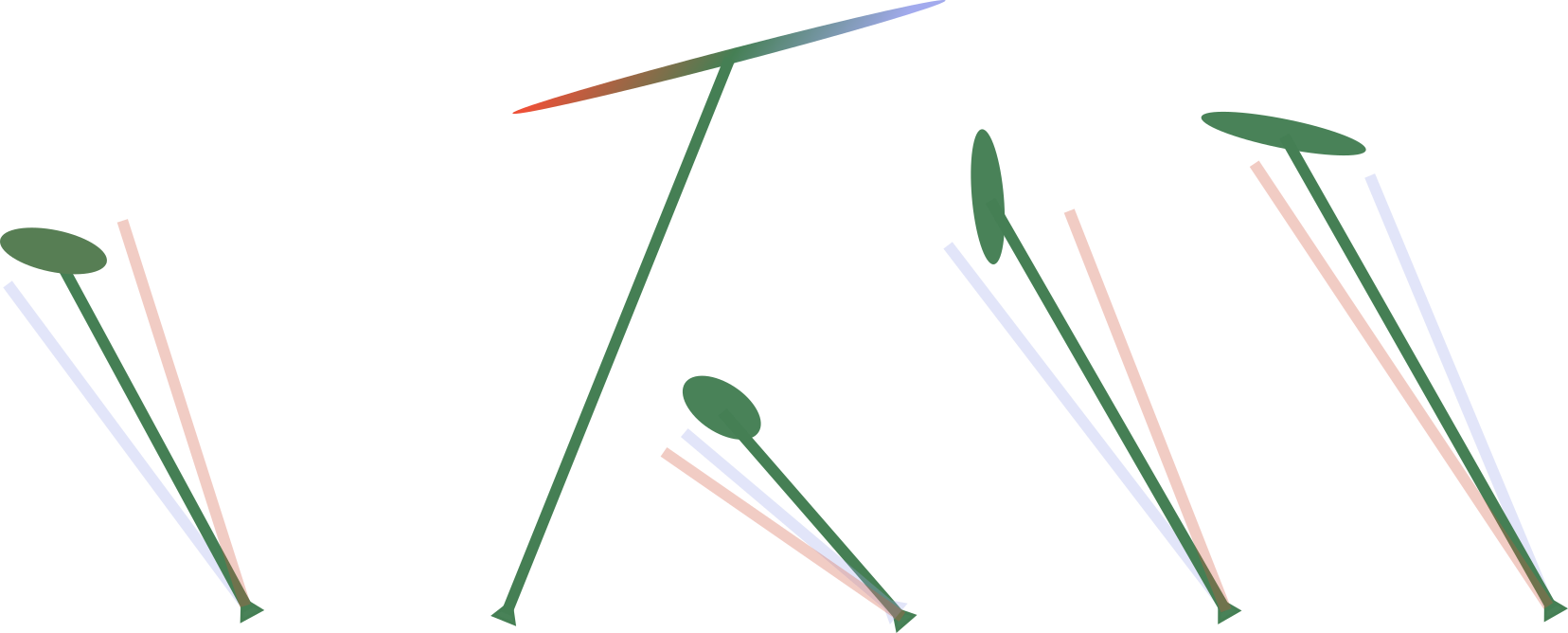}
    \caption{Cartoon of a multimode Gaussian state. The lines represent the coherent state amplitudes and the ellipses illustrate the Gaussian fluctuations. The fluctuations in the mode with largest fluctuations have a color gradient. Their beyond Bogoliubov backreaction on the coherent states results in a change of the coherent state amplitudes, that is correlated with the fluctuations (indicated by their colors), resulting in a superposition of states that have vanishing overlaps. Such a many body Schr\"odinger cat can be approximated by an incoherent mixture with very good accuracy. }
    \label{fig:sketch5}
\end{figure}

The proposed scheme is based on the modal decomposition of the Gaussian fluctuations, that can in general be written as rotated squeezed states \cite{serafini2017quantum}, schematically illustrated in Fig. \ref{fig:sketch5}.
In this representation, the quantum coherence is reflected by the reduced fluctuations in the narrow directions \cite{zurek1995decoherence}, that are canonically conjugate to the directions with large fluctuations. 
Beyond-Bogoliubov corrections, that describe the backreaction of the fluctuations on the classical field, will be argued to lead to the `measurement' of the quantum fluctuations by the classical fields, as illustrated by the purple and orange lines in Fig. \ref{fig:sketch5}. The coherence of the quantum fluctuations can subsequently neglected and the system can be represented as a classical mixture of states with much smaller fluctuations, for whose time evolution the Bogoliubov approximation is again valid.
This leads to the favorable situation where the breakdown of the Bogoliubov approximation is accompanied by the decoherence needed to save it. As it will be shown below, in the mean field limit, there is a large time window where the Bogoliubov approximation is still very good, but where decoherence has already taken place. The existence of this time window guarantees the controlled nature of the approach.

I believe that my analysis of the dynamics and intrinsic decoherence in weakly interacting Bose gases carries a fundamental new insight in closed quantum systems. 
It has been recently shown that Gaussian fluctuations can be added to Gutzwiller variational states  \cite{caleffi2020quantum} and there does not seem to be a fundamental reason why the addition of Gaussian fluctuations could not be done for other types of variational states, such as tensor network states. The present analysis would then generalize to all systems whose ground state correlations \footnote{Here, I assume that the ground state gives a good indication of the essential quantum correlations that are needed to describe dynamical states. This issue deserves further study.} are captured by a variational state.
The dynamics of almost all quantum systems would then reduce to stochastic classical motion supplemented with small quantum fluctuations.

The paper is organized as follows. In Sec. \ref{sec:ingred}, I introduce all the ingredients on which the method is based. First, in Sec. \ref{sec:singmode}, the gist of the method is illustrated for the simple example of a single mode system that evolves under a quadratic squeezing hamiltonian. In Sec. \ref{sec:tdvp}, the time dependent variational principle (TDVP) with Gaussian states is recapitulated and in Sec. \ref{sec:bog}, the squeezing under Bogoliubov time evolution and its connections to the Lyapunov spectrum are reviewed. In Sec. \ref{sec:apprclas}, the approximation by a classical mixture is explained for multimode systems and in Sec. \ref{sec:decoh}, this approximation is justified with a decoherence argument. 
Further discussion and relation to other approaches can be found in Sec. \ref{sec:discuss}. Conclusions and an outlook are given in Sec. \ref{sec:concl}.
Appendix \ref{ap:norm} gives some more details on the normal mode decomposition of Gaussian states.

\section{Ingredients \label{sec:ingred}}

\subsection{Gaussian trajectory approximation to single mode squeezing
\label{sec:singmode}}

\begin{figure*}
    \centering
    \includegraphics[width=1.\textwidth]{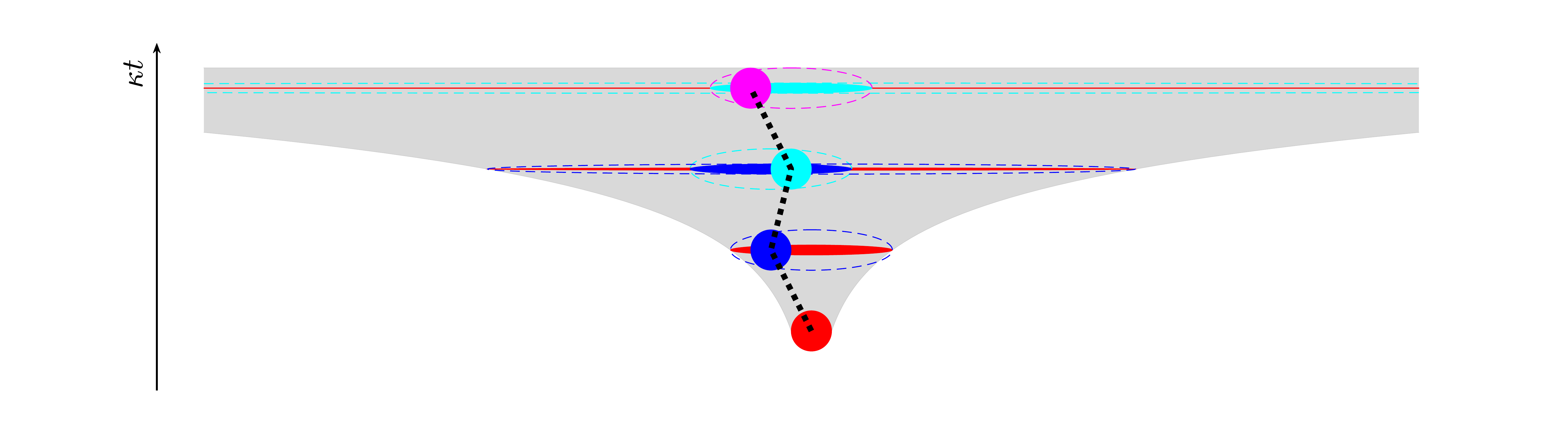}
    \caption{The original coherent state (red disk) is squeezed by the time evolution into a squeeze state (red ovals). The squeezed state is broadened (blue dashed ellipse) in order to be able to sample it with coherent states. The blue circle gives one possible realization and time evolution is continued from this state. The sampling procedure is repeated, leading to a stochastic trajectory for the mean field amplitude (black dotted line). This procedure preserves the statistics in the elongated direction, but increases the uncertainty in the narrow direction. Objects with the same style at different times are related by the Hamiltonian evolution. The figure was cropped to improve the visibility.}
    \label{fig:sketch}
\end{figure*}

The main idea upon which the approximation of a quantum system by a classical mixture is based, can be illustrated for the example of a single mode that starts in a coherent state and evolves under the `squeezing' Hamiltonian \cite{gardiner2000quantum}
\begin{equation}
    \hat H_{sq} = -i \frac{\kappa}{2} (\hat a \hat a - \hat a^\dag \hat a^\dag).
    \label{eq:Hamsqueeze}
\end{equation}
This Hamiltonian keeps the coherent amplitude constant and yields  time dependent fluctuations that read in terms of the quadratures  $\hat x=(\hat a+ \hat a^\dag)/\sqrt{2}$ and $\hat p=-i( \hat a- \hat a^\dag)/\sqrt{2}$
\begin{align}
   \sigma_x^2(t) &\equiv \langle \hat x^2 \rangle_t =\frac{1}{2}  e^{\kappa t}, \label{eq:sigmax} \\
    \sigma_p^2(t) &\equiv \langle \hat p^2 \rangle_t = \frac{1}{2} e^{-\kappa t}. \label{eq:sigmap}
\end{align}
The effect of the Hamiltonian \eqref{eq:Hamsqueeze} is to stretch the state along the $x$-direction and to squeeze it in the $p$-direction, as schematically represented by the red ellipses in Fig. \ref{fig:sketch}.
The preservation of purity under unitary evolution is reflected by the preservation of the minimal phase space volume $\sigma_x \sigma_p$ of the squeezed state, a property that is directly connected to the conservation of the phase space volume under classical Hamiltonian evolution. Starting out with equal uncertainty in the two conjugate variables $x$ and $p$, both quantum-mechanically and classically, a larger uncertainty in the $x$ variable leads to a reduction in the uncertainty in the conjugate $p$ variable.

As explained in the introduction, we wish to approximate the pure quantum state with large fluctuations by a classical mixture of states with smaller fluctuations. To illustrate the procedure, we will approximate it by a mixture of coherent states
\begin{equation}
  \hat \rho_{\rm m}=  \int d \alpha_r\, d\alpha_i \; P(\alpha_r,\alpha_i)
  |\alpha \rangle \langle \alpha|.
  \label{eq:Prep}
\end{equation}
Here, the states $|\alpha \rangle$ are coherent states  with $\alpha = \alpha_r + i \alpha_i$ and the function $P$ is the Glauber–Sudarshan $P$-distribution \cite{gardiner2000quantum}. 
Gaussian states for which a regular $P$-distribution exist are called classical, because they can be written as a classical mixture of coherent states.

Squeezed state however, are not classical, forcing us to modify the state for the decomposition \eqref{eq:Prep} to exist, i.e. to \textit{classicize} the state. 
The condition for the existence of a regular $P$-distribution being that both the $x$ and $p$ variances should be larger than one \cite{serafini2017quantum}, the minimal modification is
\begin{equation}
\sigma_p \rightarrow 1.
\label{eq:mod_sigma_p}
\end{equation}
The $P$-distribution for this classicized state is then
\begin{equation}
    P^{(2)}(\alpha_r,\alpha_i) = \frac{1}{\sqrt{\pi (\sigma_x^2-1)}}
    \exp\left[-\frac{ \alpha_r^2} {\sigma_x^2-1}\right] \delta(\alpha_i).
\end{equation}
The price that we have to pay in order to represent our state as a classical mixture of coherent states  is that we lose the reduction of the variance in the narrow direction, as graphically illustrated in Fig. \ref{fig:sketch} by the difference between the red filled and the blue dashed ellipses.

At first sight, the integral in Eq. \eqref{eq:Prep} could seem to form an obstacle, especially when several consecutive classical approximations are made. In practice, the integration over all coherent state amplitudes in \eqref{eq:Prep} can be done with a Monte Carlo sampling. Such \textit{unraveling} is routinely done in the quantum trajectory method for open quantum systems \cite{breuer2002theory} and is schematically represented in Fig. \ref{fig:sketch} by the randomly selected colored circles. The resulting trajectory is represented by the dashed black line.

In order to make the connection to decoherence and to the standard quantum trajectory method \cite{breuer2002theory} even more direct, the approximate decomposition of the state \eqref{eq:Prep} can also be achieved by considering the quantum trajectories that follow from dissipative dynamics with $\hat x$ as jump operator.
This dissipation preserves the Gaussianity of the state and the first two moments then obey the following stochastic equations:
\begin{align}
    d\langle \hat x \rangle &= \sqrt{\gamma} \langle \delta \hat x^2 \rangle dW \\
    d\langle \hat p \rangle &=0 \\
    d \langle \delta \hat x^2 \rangle  &= 
    -\gamma \langle \delta \hat x^2 \rangle^2 dt \\    
     d\langle \delta \hat p^2 \rangle &= \gamma dt
\end{align}
where $\delta \hat x = \hat x-\langle \hat x \rangle$ and $dW$ is Gaussian white noise with variance $\langle dW_t dW_{t'}  \rangle  =\delta_{t,t'}\; dt$.

After an evolution such that $\langle \delta \hat p^2 \rangle = 1/2$, one obtains in the limit $\langle \delta \hat x^2 \rangle_0\gg 1$ that the variance in $\langle \hat x \rangle$ due to its stochastic evolution equals $\langle \delta \hat x^2 \rangle_0$: the quantum fluctuations are then  converted into classical fluctuations in the same way as through the modification \eqref{eq:mod_sigma_p} followed by the decomposition in terms of the $P$-distribution.
This shows that our approximation and subsequent decomposition in the previous section actually corresponds to a quantum trajectory for the system with dissipation in the $x$-direction, the direction with the largest fluctuations.

\subsection{TDVP with Gaussian states \label{sec:tdvp}}

For the quadratic Hamiltonian, the solution (\ref{eq:sigmax}), (\ref{eq:sigmap}) is exact and the rewriting \eqref{eq:Prep} is not useful. However when beyond quadratic terms are added to the Hamiltonian, the Gaussian approximation can be much more accurate after the decomposition \eqref{eq:Prep} than the ordinary time dependent Bogoliubov theory. Before turning to this discussion, we first recapitulate the time dependent variation principle (TDVP) with Gaussian states \cite{boudjemaa2010variational, verstraelen2020gaussian, guaita2019gaussian}.

For the quadratic Hamiltonian in the previous section, the dynamics remains in the manifold of Gaussian states, but for nonlinear Hamiltonians the Gaussianity can be preserved only by projecting it back into the Gaussian manifold within the framework of the TDVP \cite{kramer1981lecture}.
Since Gaussian states are fully characterized by their first and second moments ($\alpha_i=\langle \hat a_i \rangle$ and $\sigma^{(a)}_{ij}$),
the TDVP with Gaussian states reduces to the equations of motion for those moments, where Wick's theorem is used in order to factorize correlators of higher than second order. 

To be specific, we consider the Bose-Hubbard model with nearest neighbor tunneling, described by the Hamiltonian
\begin{equation}
    \hat H = -J \sum_{ \langle i,j \rangle } (\hat a^\dag_j \hat a_i + \hat a^\dag_i \hat a_j ) 
    +\frac{U}{2} \sum_i \hat a_i^\dag  \hat a_i^\dag  \hat a_i  \hat a_i,
\end{equation}
 where $J$ is the hopping rate, $\langle i,j \rangle$ runs over nearest neighbours and $U$ is the interaction strength.

The extended GPE for the coherent amplitudes $\alpha_n=\langle \hat a_n \rangle$ reads
\begin{equation}
    i\frac{\partial \alpha_n}{\partial t} = -J \sum_{n'} \alpha_n' + U |\alpha_n|^2 \alpha_n + U(\alpha_n n_{nn}+\alpha^*_i c_{nn}),
    \label{eq:bh_GPE}
\end{equation}
where $n_{nm} = \langle \delta \hat a_n^\dag \delta \hat a_m\rangle $ and $c_{nm} = \langle \delta \hat a_n \delta \hat a_m \rangle$.
and the sum over $n'$ runs over the nearest neighbors of $n$. The first two terms are the ones from the usual GPE and the last one describes the correction due to the quantum fluctuations. These in turn are governed obey the following equations of motion
\begin{align}
    i \frac{\partial n_{nm}}{\partial t} &= -J \sum_{n'} (n_{n m'}-n_{n' m}) \nonumber \\
    &+2Un_{nm}(|\alpha_n|^2-|\alpha_m|^2 + n_{nn}-n_{mm}) \nonumber \\
    &+Uc_{nm}(\alpha_n^{*2}+c_{nn}^*)
    -U c_{nm}^*(\alpha_m^2+c_{mm}), \label{eq:mot_n} \\
    i \frac{\partial c_{nm}}{\partial t} &= -J \sum_{n'} (c_{n m'}+c_{n' m}) \nonumber \\
    &+2Uc_{nm}(|\alpha_n|^2-|\alpha_m|^2 + n_{nn}+n_{mm}) \nonumber \\
    &+U n_{nm}(\alpha_n^{2}+c_{nn})
    + U c_{mn}(\alpha_m^2+c_{mm}) \nonumber \\
    &+\delta_{n,m}(\alpha_n\alpha_m+c_{nm}). \label{eq:mot_c}
\end{align}

The last term in Eq. \eqref{eq:mot_c} originates from the canonical commutation relation and captures the effects of quantum fluctuations. In terms of quasi-particle scattering, it is responsible for the spontaneous scattering processes where both final states are empty, the crucial process that is missing from the classical field theory \cite{griffin1996bose}.
It is readily seen from the Boltzmann equation that spontaneous scattering is essential to avoid the UV catastrophe of classical field theory.

The terms in Eqs. \eqref{eq:bh_GPE}-\eqref{eq:mot_c} that are nonlinear in the quantum fluctuations do not give a systematic expansion in terms of the small parameter of the system \cite{castin1998low} and will be neglected. This is equivalent to considering linearized Heisenberg equations of motion for the fluctuation operators
\begin{equation}
    i \frac{\partial}{\partial t}  \delta\! \hat A = B(t) \, \delta \!  \hat A,
    \label{eq:bogform}
\end{equation}
where $ \mathbf{\hat A}=(\hat a_1, \hat a_2, \ldots,\hat a_1^\dag, \hat a_2^\dag,\ldots)$.
The Bogoliubov evolution matrix $B$ is obtained from the Hessian of the classical Hamiltonian as
\begin{equation}
    B_{ij}(t) = \bar J\; \frac{\partial \mathcal{H}[\boldsymbol{\alpha}(t),\boldsymbol{\alpha^*}(t)]}{\partial A_i \partial A_j},
\end{equation}
where  $\mathcal{H}(\boldsymbol{\alpha},\boldsymbol{\alpha^*}) = \langle \boldsymbol{\alpha^*} | \hat H |\boldsymbol{\alpha}\rangle $ 
and $\bar J$ is the symplectic form in the amplitude basis:
\begin{equation}
     \bar J = \begin{pmatrix}
         \mathds{1}  & 0 \\ 0 & -\mathds{1} 
    \end{pmatrix}.
    \label{eq:Jdef}
\end{equation}

\subsection{Squeezing under Bogoliubov time evolution \label{sec:bog}}

In Sec. \ref{sec:singmode}, we have looked into the Gaussian trajectory approximation for a single mode system. Our approximation consisted of the broadening in the squeezed direction, that allowed us approximate the squeezed state by a mixture of states with much smaller fluctuations. 
For the analysis of the fluctuation modes of the system, it is useful to construct the correlation matrix \cite{serafini2017quantum}
\begin{equation}
    \sigma_{ij}^{(a)} = \langle \{ \delta A_i , \delta A^\dag_j \} \rangle,
\end{equation}
where the vector $\mathbf{\hat A}$ is defined as $\mathbf{\hat A} = (\hat a_1, \hat a_2, \ldots, \hat a_1^\dag, \hat a_2^\dag, \ldots)^T$
and the curly brackets represent the anticommutator. In terms of the normal and anomalous correlation matrices $n_{ij} = \langle \delta\hat a_i^\dag \delta a_j\rangle$ and $c_{ij} = \langle \delta \hat a_i \delta a_j\rangle$,
the correlation matrix has the block form
\begin{equation}
    \sigma^{(a)} = \begin{pmatrix}
        2 n + \mathds{1} & 2c \\ 2c^* & 2n + \mathds{1}
    \end{pmatrix}
\label{eq:sigmablock}
\end{equation}

Let us now consider the case where we start from a coherent state that is evolved with the Gaussian TDVP. 
The Bogoliubov equation \eqref{eq:bogform} for the fluctuations can be solved formally as
\begin{equation}
    \delta\! \hat A (t) = U_B(t)  \, \delta\! \hat A (0),
\end{equation}
where the evolution matrix is the solution of
\begin{equation}
    i \frac{\partial}{\partial t} U_B(t) = B(t)\, U_B(t)
\end{equation}
with initial condition $U_B(0)=\mathds{1}$.
The matrix $U_B(t)$ represents a symplectic transformation \cite{serafini2017quantum} and therefore satisfies $U_B^\dag(t)\, \bar J \, U_B(t) = \bar J$. 

In terms of $U_B$, the time dependence of the correlation matrix is
\begin{equation}
    \sigma^{(a)}(t) = U_B(t) \sigma^{(a)}_0 U_B^\dag(t).
\end{equation}
When starting from a coherent state, $\sigma^{(a)}_0=\mathds{1}$, and the time dependent correlations then reduce to $\sigma^{(a)}(t) =  U_B(t)  U_B^\dag(t)$.

Since the matrix $U_B$ describes the linear fluctuations on top of the classical Gross-Pitaevskii solution, the matrix $U_B(t) U_B^\dag(t)$ is directly related to the (finite time) Lyapunov exponents of the GPE, that are defined as \cite{ott2002chaos}
\begin{equation}
    \lambda_n(t) = \left\{ \frac{1}{t} \log[d_n(t)] \right\}^{1/2},
\end{equation}
where the $d_n(t)$ are the eigenvalues of the matrix $U_B(t)U_B^\dag(t)$.

As a consequence of the symplectic structure of $U_B$, the eigenvalues of the matrix $U_B U_B^\dag$ come in pairs $(d_i,1/d_i)$, which results in pairs of Lyapunov exponents with opposite signs.
Hence, the phenomenology from Sec.~\ref{sec:singmode} is carried over to the multimode case: Hamiltonian time evolution results in the amplification of fluctuations in certain directions and squeezing in the conjugate ones. 

The single mode example illustrated that it is the small uncertainty in the squeezed direction that prevents the representation of the state as a classical mixture of coherent states.
In dynamical terms, it are thus the negative Lyapunov exponents that express the quantum coherence in the system.
But the fact that the quantumness originates from exponentially contracting variables implies that the coherence of the quantum superposition is expressed by a relatively small contribution to the correlation matrix. It can therefore hardly be expected to be essential for the further evolution of the system. 

\section{Classicization and unraveling of multimode Gaussian states \label{sec:apprclas}}

Large fluctuations in a strongly squeezed multimode Gaussian state can again be eliminated along the same lines as in the single-mode case in sec. \ref{sec:singmode}. In order to carry out this procedure, it is useful to first write the Gaussian state in its normal mode form \cite{serafini2017quantum}. To this purpose, one rewrites the correlations in terms of the quadrature operators $\hat x_i=(\hat a_i + \hat a_i^\dag)/\sqrt{2}$ and $\hat p_i=-i(\hat a_i - \hat a_i^\dag)/\sqrt{2}$:
\begin{equation}
     \mathbf{\hat A} = \bar U \mathbf{\hat r}
\end{equation}
with $ \mathbf{\hat r} = (\hat x_1,\hat x_2,\ldots,\hat p_1,\hat p_2, \ldots  )^T$ and
\begin{equation}
    \bar U = \frac{1}{\sqrt{2}} \begin{pmatrix}
        \mathds{1} &   i \mathds{1}\\
        \mathds{1} & - i \mathds{1}
    \end{pmatrix}.
\end{equation}
In the quadrature variables, the correlation matrix is
\begin{equation}
    \sigma^{(xp)}_{ij} =  \langle \{ \delta  \hat r_i, \delta  \hat r_j^\dag \} \rangle =\bar U\sigma^{(a)} \bar U^\dag
    \label{eq:cortf}
\end{equation}
and for a pure state, it can be written as
\begin{equation}
    \sigma^{(xp)} = O_1 \Sigma^2 O_1^T.
    \label{eq:sigmaxpdiag}
\end{equation}
The diagonal matrix $\Sigma^2$ contains the eigenvalues, that come in pairs $z_i^2$ and $1/z_i^2$ (see appendix \ref{ap:norm} for more details).  
The transformation matrix $O_1$ is both orthonormal and symplectic   and specifies in which directions the fluctuations are stretched or compressed. Note that these modes are selected by the dynamics itself and there is no need to determine beforehand a preferential basis.

Following the procedure from the single mode case in Sec. \ref{sec:singmode}, the Gaussian state can now be approximated by a mixture.
In order to remain as faithful as possible to the original state, one may wish to retain some squeezing. To this aim, the squeezing matrix is written as
\begin{equation}
    \Sigma  = \Sigma^{(1)} \Sigma^{(2)},
    \label{eq:sigma_dec}
\end{equation}
where the (not too large) squeezing in $\Sigma^{(1)}$ will be kept and where we will approximate the state described by $\Sigma^{(2)}$ by a classical mixture of coherent states. Note that one may opt to put part of the squeezing of a certain mode in $\Sigma^{(1)}$ and another part in $\Sigma^{(2)}$ in order to reduce the squeezing rather than fully eliminate it. In the graphical representation of Fig. \ref{fig:sketch}, this means that one samples with ellipses rather than with circles.

In analogy with the single mode case, the criterion for a Gaussian state to be classical is that all the eigenvalues of its correlation matrix are larger than one. This is clearly not the case for a squeezed pure state, whose eigenvalues come in pairs $(z^2_i,1/z^2_i)$. But by the modification $\Sigma^{(2)} \rightarrow \tilde \Sigma^{(2)}$, obtained by the replacement
\begin{equation}
1/z^{(2)}_i \rightarrow 1/\tilde z^{(2)}_i = 1 ,   
\label{eq:neglect_sq}
\end{equation}
the $\Sigma^{(2)}$-state is classicized. It can then be unraveled as
\begin{equation}
    \hat \rho_{cl}(\Sigma^{(2)}) = \int d \boldsymbol{\alpha}  \; P_2(\boldsymbol{\alpha}) \;
    |\boldsymbol{\alpha}\rangle \langle \boldsymbol{\alpha} |,
    \label{eq:rhocldec}
\end{equation}
with
\begin{equation}
    P_2(\boldsymbol{\alpha}) = \prod_n \frac{1}{\sqrt{\pi \tilde \sigma_n^2}}
    \exp\left[-\frac{ \alpha^2_{r,n}} {\tilde \sigma_n^2}\right] \delta(\alpha_{i,n}),
\end{equation}
where $\tilde \sigma_n^2 = (z^{(2)}_n)^2-1$.
As explained in the single mode case, the integral over the classical fluctuations can be done numerically by Monte Carlo sampling as in the quantum trajectory method.

The result of the classicization and unraveling procedure is then
\begin{align}
    \boldsymbol{\alpha} &\rightarrow \boldsymbol{\alpha} 
    + \sum_n  \tilde \sigma_n \xi_n\;  \bar U O_1 \Sigma^{(1)} \boldsymbol{x}_n, 
    \label{eq:dalpha_stoch} \\
    \sigma^{(a)} &\rightarrow \bar U^\dag O_1 \, (\Sigma^{(1)})^2 \, O_1^T \bar U.
    \label{eq:sigma_red}
\end{align}
Here, the summation runs over all modes whose squeezing is eliminated, $\xi_n$ is a real Gaussian random number with zero mean and unit variance and the vector $\boldsymbol{x}_n$ contains a single one at its $n$th element and zeros otherwise.

\section{Decoherence of the quantum fluctuations by the classical field \label{sec:decoh}}


We now come to the crux of the justification of the proposed Gaussian trajectory method. The physical content of the classicization approximation is the neglect of the coherence between fluctuations. The reward is that nonlinear corrections to the Gaussian theory can be kept small at all times. As we will show in this section, these two aspects become compatible in the mean field limit.

The reason why the Gaussian theory breaks down can be most easily understood by considering the time evolution of the different trajectories after classicization and unraveling as discussed in the previous section. For each realization, the coherent fields $\alpha_i$ will undergo a different evolution. This is in contrast to the evolution within the purely Gaussian approximation, where there is a single classical field. Physically, this means that the Gaussian theory misses the backreaction of the quantum fluctuations on the classical fields.

In terms of correlation functions, this backreaction would be captured to by the third order correlators \cite{van2018prethermalization,colussi2020cumulant}, but these in turn give rise to fourth order ones and so on. In a correlation function approach to a classical statistical problem, this would correspond to trying to approximate the probability distribution by doing a series expansion of its characteristic function. In general however, there is no guarantee that this expansion has good convergence properties, rather the contrary \cite{Marcinkiewicz1939}. A much more efficient approach is to sample the probability distribution as it is done in molecular dynamics or Metropolis-Hastings Monte Carlo simulations. This is precisely the strategy that I suggest here for the quantum case. This approach is only valid when the coherence between different realizations of the fluctuations is small. I will show below that the backreaction of the fluctuation on the classical field can be seen as a measurement of the fluctuation by the classical field, leading to the decoherence of the quantum fluctuations and giving consistency to the method.

In order to keep the mathematics for the analysis as simple as possible (but not simpler), let us consider a classical field $\alpha_i$ that is perturbed by a classical fluctuation $\delta_i(t)$, and consider the back-reaction of this fluctuation on the field $\alpha_i$. One can think about these fluctuations as sampling the Wigner distribution. The fluctuations then evolve to first approximation under the Bogoliubov equation \eqref{eq:bogform} from an initial value, that lies along a direction stretched by the dynamics and has a magnitude $\delta_0$ that is of the order of the corresponding element of the matrix $\Sigma$ in Eq. \eqref{eq:sigmaxpdiag}.
The backreaction of the fluctuation $\delta_i$ then modifies the dynamics of the $\alpha_i$-field:
\begin{equation}
    i\frac{\partial  \alpha'_i}{\partial t} = -J (\Delta \alpha')_i + U |\alpha'_i|^2\alpha'_i + 2 U |\alpha'_i|^2 \delta_i + U {\alpha'}^2_i \delta_i^*,
    \label{eq:beybog}
\end{equation}
where the prime indicates the presence of the fluctuation. Note the difference between Eqs. \eqref{eq:bh_GPE} and \eqref{eq:beybog}, in particular the quadratic versus linear scaling with the fluctuation amplitude.
After evolution over some time $t$, there will be a difference between the solution $\alpha_i(t)$ of the GPE without the perturbation and the $\alpha'_i(t)$ in the presence of the fluctuation. The Bogoliubov approximation breaks down when this difference becomes too large. In terms of correlation functions, this means that the third order cumulants can no longer be neglected \cite{van2018prethermalization,colussi2020cumulant}. 

Let us now look at the above discussion in terms of decoherence. Due to its effect on the classical field $\alpha'_i$, different values of the fluctuation $\delta_0$ will induce a loss of overlap between the correspondingly different fields $\alpha'_i$ and $\alpha_i$.
The resolution of the measurement of the fluctuation by the classical field is quantified by the overlap
\begin{equation}
    |\langle \boldsymbol{\alpha}'(t) | \boldsymbol{\alpha} (t)\rangle|^2 
    = e^{-\sum_i |\alpha'_i(t)-\alpha_i(t)|^2 } \approx e^{-|\delta_0|^2 /2 \sigma_\delta^2(t)}, \label{eq:decoh_alpha}
\end{equation}
where the last form highlights the (in leading order) linear scaling of $|\alpha'_i-\alpha_i|$ with $\delta_0$.
Physically, a small overlap between the states $\alpha_i$ and $\alpha'_i$ implies that the fluctuation has been `detected' by the classical field.

For a classical field with many degrees of freedom, it is safe to neglect the possibility of a revival and for all practical purposes one can neglect any future coherence after its initial decay. The quantity $\sigma_\delta^2(t)$ can then be interpreted as the resolution with which the classical field measures the fluctuation $\delta_0$. 
We thus conclude that the backreaction of a quantum fluctuation on all the other classical modes leads to the decoherence of the quantum fluctuation. The classical field plays here the role of the classical `apparatus' in the usual Copenhagen interpretation of quantum mechanics and there is no need for any further decoherence by an environment.

We now have a prescription for when to apply the classicization as a compromise between keeping the Bogoliubov theory valid and preserving as much as possible the quantum coherence of the fluctuations.
Conveniently, for the weakly interacting Bose gas the breakdown of the Bogoliubov approximation is determined by the relative difference of $|\alpha'-\alpha|$ with respect to $\sqrt{\bar n}$, while the resolution $\sigma_\delta$ is determined by the absolute difference. In the mean field limit, where $\bar n$ tends to infinity, the classicization can therefore be safely applied in the regime where the Bogoliubov approximation is still accurate, with a substantial freedom for choosing the time when the classicization is performed. Thanks to the exponential scaling of the overlap \eqref{eq:decoh_alpha}, the method becomes exponentially accurate for increasing mean field parameter $\bar n$  (keeping $U\bar n$ fixed), that plays here the role of an inverse Planck constant.

In summary, the proposed method consists of the variational time evolution with Gaussian states according to Eqs. \eqref{eq:bh_GPE}, \eqref{eq:mot_n} and \eqref{eq:mot_c}, interrupted by classicization stages. During the dynamics, squeezing  develops along certain directions that can be determined by diagonalization of the correlation matrix $\sigma^{(xp)}$, cf. Eq. \eqref{eq:sigmaxpdiag}.
The decoherence of these fluctuations can be monitored by perturbing the GPE as in Eq. \eqref{eq:beybog} with for the initial states of $\delta_i$ the directions of the largest fluctuations and with magnitude their squeezing parameters. When the overlap with the unperturbed GPE solution becomes small compared to some threshold, the quantum fluctuation can be classicized according to Eqs. \eqref{eq:dalpha_stoch} and \eqref{eq:sigma_red}.

\section{Discussion \label{sec:discuss}}

\subsection{Generalization to other variational states}

The Gaussian trajectory approach to the dynamics of the weakly interacting Bose gas is based on coherent states being a decent approximation to the ground state. For other systems, such as strongly interacting Bose gases in optical lattices, coherent states are not adequate, but other variational states like Gutzwiller states can be used instead. It has been recently demonstrated that Gaussian fluctuations can be added to Gutzwiller variational states \cite{caleffi2020quantum}. In this approach, the TDVP with the Gutzwiller states is seen as the classical theory that is quantized by replacing the variational parameters by $c_i$ by $c_i +\delta \hat c_i$. Hereafter, the Hamiltonian is expanded up to quadratic order in the quantum fluctuations $\delta \hat c_i$. The squeezing in the ground state turns out to be limited, expressing that the Gutzwiller state is already a decent approximation to the ground state.
It can be expected that the addition of Gaussian quantum fluctuations not only works for coherent and Gutzwiller states, but for any variational ansatz, where Gaussian fluctuations then capture contributions from the double tangent space \cite{haegeman2013post}. A good variational ansatz then always turns a quantum system into a weakly interacting (multicomponent) Bose gas \footnote{When there are fermionic degrees of freedom, the fluctuations will be both of fermionic and bosonic nature \cite{calefficollective2022}}.

The TDVP dynamics of the variational parameters being nonlinear, it is in general chaotic, which implies an exponential growth in time of the quantum fluctuations. Even when a variational approximation is able to capture ground state properties excellently, it will break down for time evolution, as it is indeed well known for MPS \cite{schollwock2011density}, for which the connection between classical chaos and the breakdown of the accuracy of MPS for time evolution was pointed in Ref. \cite{hallam_lyapunov_2019}.

When the Bogoliubov theory on top of the variational wave function is weakly interacting, the Gaussian trajectory method becomes applicable. 
As such, the Gaussian trajectory method should be a universal ingredient for describing the dynamics of closed quantum systems. In this scheme, the role of the variational state is to take care of the short range `high energy' correlations, where the remaining fluctuations at larger scales can be described by the Gaussian trajectories.

\subsection{The Wigner distribution perspective \label{subsec:wigner}}

The Wigner distribution, especially in the truncated Wigner approximation (TWA), is a widely used tool for the study of the weakly interacting Bose gas \cite{sinatra2002truncated,polkovnikov2010phase}. In one dimension, where the UV catastrophe is less severe, it leads to good quantitative predictions \cite{polkovnikov2008breakdown} and in higher dimensions it can be used for short times \cite{sinatra2002truncated} as well as for systems with small fluctuations \cite{carusotto2008numerical}.

The problem in nonlinear systems is that the exact Wigner distribution develops interference fringes when the phase space distribution becomes tightly squeezed \cite{zurek2003decoherence}, a feature that is characteristic for Schr\"odinger cat states \cite{raimond2006exploring}.
Those intricate interference fringes make the Wigner distribution very hard to compute (as difficult as to solve the full many body problem) and it is hard to avoid making systematic errors, even though some schemes to go beyond the TWA have been developed \cite{polkovnikov2010phase,sels2014variational}. 

Dissipative systems are easier to describe, because dissipation washes out the interference fringes \cite{zurek2003decoherence} and the TWA can be used for long times \cite{carusotto2013quantum}.
In the presence of a dissipation with jump operator $\gamma \hat x$, the Wigner function dynamics gets an additional diffusion term:
\begin{equation}
    \frac{\partial W }{\partial t} = \frac{\gamma}{2}  \frac{\partial^2 W }{\partial p^2},
\end{equation}
It broadens the Wigner distribution so that it counteracts the development of interference fringes, and therefore improves the simulability of the system. 

In this work, it is proposed to complement the Hamiltonian dynamics with quantum jumps corresponding to the measurement of the observables corresponding to the most elongated directions in the phase space distribution. This corresponds to adding a diffusion term to the TWA in the directions with the tightest squeezing.

Previous works have made the relation between the occurrence of fringes in the Wigner distribution and the quantum to classical transition \cite{zurek2003decoherence}. What I suggest in this work is to add the dissipation to the dynamics before the fringes have developed. 
For the disturbance of the system due to the added dissipation, one can make the same arguments as in the discussion of the Gaussian approximation and one therefore expects a window for dissipation strength to exist where it keeps the system classical while not significantly disturbing the dynamics.

In order to compute the directions in which to add the dissipation, one has to compute the time evolution of the linear fluctuations. The computational cost of computing the dynamics in the tangent space along the classical trajectory being the same as the evolution of the correlation matrix, it is expected that the numerical complexity of the TWA complemented with the measurements is the same as the Gaussian trajectory method.

\subsection{Entropy growth}

For the case of Gaussian states evolving under quadratic Hamiltonians, it was shown by Bianchi {\em et al.} \cite{bianchi2018linear} that the entanglement entropy growth rate of a subsystem $A$ with $N_A$ sites is given by the sum of the $2N_A$ largest Lyapunov exponents:
\begin{equation}
    \frac{d S_{A}}{dt}  = \sum_{i=1}^{2N_A} \lambda_i,
    \label{eq:entr_rate_quadr}
\end{equation}
where the Lyapunov exponents are ordered in decreasing order ($\lambda_1 \geq \lambda_2 \geq \ldots$).
For a subsystem within the range $N_I \leq 2N_A \leq 2N-N_I$, where $N_I$ is the number of nonvanishing Lyapunov exponents and $N$ is the total number of lattic sites, the growth rate coincides with the Kolmogorov-Sinai entropy rate, that is according to Pesin's theorem  equal to the sum of the positive Lyapunov exponents \cite{ott2002chaos}.

In the present Gaussian trajectory method the initial entanglement entropy is governed by the quadratic Bogoliubov approximation and therefore grows  at the rate \eqref{eq:entr_rate_quadr}, where the Lyapunov exponents have to be interpreted as finite time exponents. We do not let it grow much however, but rather convert it into von Neumann entropy.
The production of von Neumann entropy in chaotic quantum systems that are subject to decoherence was already discussed in the pioneering work by Zurek and Paz \cite{zurek1995decoherence}. They argue that, in the limit of weak decoherence, its growth tends to the Kolmogorov-Sinai entropy rate. The Zurek-Paz argument also holds for the present approach, illustrating that entanglement entropy is indeed converted into von Neumann entropy.

The important difference between the present approach and the Zurek-Paz setting is that they introduced decoherence from the outside, where it originates here from within the system itself. This self decoherence is only possible for systems with a large number of degrees of freedom. Indeed, for systems with a small number of degrees of freedom, it cannot be excluded that they `forget' the result of a measurement, i.e. that they end up in the same position in phase space, irrespective of the outcome of a past measurement. According to the usual arguments form statistical physics \cite{huang2008statistical}, this becomes highly unlikely for systems with a large number of degrees of freedom.

\subsection{Thermalization in closed quantum systems}

Spurred by experiments on systems of ultracold atomic gases, considerable efforts have been invested in the elucidation of the issue of thermalization in closed quantum systems under unitary evolution \cite{ueda2020quantum}. One of the conceptual ways to understand the approach to thermal equilibrium is through the so-called eigenstate thermalization hypothesis (ETH) \cite{d2016quantum}. This conjecture about the nature of the eigenstates of nonintegrable quantum systems allows to reduce thermalization to dephasing. 
This picture stands in stark contrast to that of classical physics, where thermalization takes place by the exploration of phase space.
While conceptually enlightening, the ETH does not really bring a practical method for the efficient computation of quantum dynamics. Numerical verification of the conjecture could actually only be done when sufficient computing power became available \cite{rigol2008thermalization}. The reason is that the exact eigenstates of a many body system are highly entangled states for which in general no efficient representation is known. 

Thermalization in closed systems has also been related to the formation of entanglement: subsystems acquire thermal properties if they are sufficiently entangled with the rest of the system \cite{linden2009quantum,ueda2020quantum}. 
What is shown in the present work is how the entanglement entropy can be converted into classical entropy. The thermalization mechanism that follows from the Gaussian trajectory method is then much closer that of classical physics. Indeed, the classical fields explore phase space during their stochastic dynamics so that one expects that after a long evolution time, the system will explore all accessible regions in phase space.

It is also interesting to consider the work of Leviatan \textit{et al.} \cite{leviatan2017quantum} on the description of thermalization in closed systems with MPS simulations in the present context.
They found that at high enough energy, even low bond dimension states can well describe the approach to thermal equilibrium. Considering the MPS elements as classical fields, this regime is the one where classical equipartition holds and quantization is not important. However, when the energy is lowered, quantum fluctuations can no longer be neglected and the purely classical approximation starts to suffer from a UV catastrophe. It can be expected that the Gaussian trajectory approach on top of the MPS dynamics is able to cure this issue. From the point of view of entanglement, the Gaussian fluctuations are able to capture the long range entanglement that the low bond dimension MPS is missing.

The idea of adding decoherence to describe a closed quantum system's dynamics toward equilibrium has also been explored in simulations based on tensor network states (see e.g.
\cite{surace2019simulating,Hauschild2018finding,white2018quantum,rakovszky2022dissipation} and references therein).
To the best of my knowledge no use was made of quantum trajectories in this type of studies. I believe that the trajectory method could have the crucial advantage that it is able to represent classical correlations at a relatively small cost, while these still require large bond dimensions in a matrix product operator representation \cite{wellnitz2022rise}.


\subsection{The measurement problem}

The approximation of the Bose gas by coherent states is called the `classical field' approximation \cite{brewczyk2007classical}, referring to the fact that coherent states are the most classical of quantum states since they are as well localized in phase space as allowed by the Heisenberg principle.
The classical status of the coherent states is also reflected by the fact that within the TDVP their dynamics is governed by classical Hamiltonian dynamics. 
The Gaussian fluctuations on top of the coherent states are then the `quantum fluctuations', since they express enhanced uncertainty on certain observables.
When adopting the view that the variational parameters constitute a classical `system' and the Gaussian fluctuations a quantum `system', the Gaussian trajectory method may also shed light on the interpretation of quantum mechanics. In particular, it seems to give a picture that is very close to the standard Copenhagen interpretation.

This view is for example outlined by Landau and Lifshitz \cite{landau_quantum_2013}:
``The possibility of a quantitative description of the motion of an electron requires the presence also of physical objects which obey classical mechanics to a sufficient degree of accuracy. If an electron interacts  with such a `classical object', the state of the latter is, generally speaking, altered. The nature and magnitude of this change depend on the state of the electron, and therefore may serve to characterize it quantitatively.''
The measurement (and associated decoherence) is here clearly attributed to the altered state of the classical system. My discussion of the intrinsic decoherence in the weakly interacting Bose gas, where the Gaussian quantum fluctuations are measured by the classical coherent field is fully in line with this philosophy. 

It is however in some contrast to the usual approaches developed in the last decades for the weakly interacting Bose gas, where the point of view is taken that the quantum fluctuations decohere the modes in the classical field region \cite{davis_wright_gasenzer_gardiner_proukakis_2017,gardiner2000quantum,bijlsma2000condensate,zaremba1999dynamics}.
Those approaches are more in line with the decoherence picture, where the world is separated in a system of interest, a measurement apparatus and an environment \cite{zurek2003decoherence}. It is the role of the environment to make sure that any coherent superpositions of the conglomerate system-apparatus are destroyed. In the present work, I do away with the need for an environment. The reason why this is possible for the multimode Bose gas is that the classical field dynamics has so many degrees of freedom that it takes away worries for future revivals of quantum interferences with no more assumptions than the usual ones to counter Loschmidt's and Zermelo's objections to Boltzmann's analysis of the equilibration of classical systems.
The present analysis thus points to a fundamental link between the second law and quantum measurement, that is anyway present in practice: there is always entropy production in the detector when a measurement is done.

Finally, it is also in agreement with the general properties of quantum measurements that there is some freedom in placing the ``Heisenberg cut'', i.e. the exact time when the classicization is carried out. This freedom, that becomes larger when going deeper in the mean field limit, was discussed in Sec. \ref{sec:decoh}.

\subsection{Possible fundamental implications}

To the best of my knowledge, the present paper is the first to propose a constructive method to deal with the generic time evolution of a nonintegrable quantum system that scales polynomially with system size.
The emerging picture is that of chaotic classical dynamics with quantum fluctuations on top, that leads to a combination of deterministic classical dynamics supplemented with quantum jumps. I have moreover argued that a similar picture should hold for any quantum system for which a good variational description exists. 

It then becomes tantalizing to think that this may tell us something about the nature of our universe. Under the assumption that quantum mechanics is fundamental, the state of the universe should undergo unitary evolution in some Hilbert space. If a good variational description exists, a good approximation to this unitary evolution would be given by the Gaussian trajectory method. 

The most straightforward identification of the quantum jumps with reality would be the outcomes of quantum-mechanical measurements (with or without observer) and a single quantum trajectory would correspond in this picture with a single realization of the history of the universe, a single branch in the many-worlds interpretation \cite{everett1957relative}. 

On a speculative note, the classical fields would not only describe the states of macroscopic objects, but also the properties of space and time itself. 
This picture is very close to how we already describe reality by means of small fluctuations (quantum field theory) on top of a classical background (classical objects and spacetime metric).
Speculations of an emergent spacetime metric have been made in the context of the AdS-CFT correspondence \cite{van2010building,verlinde2011origin}, but a spacetime metric emerging from fluctuating classical fields  would do away with the difficulties of interpreting quantum states that consist of superpositions of macroscopically distinct spacetimes.
This speculation is somewhat supported by studies on analog Hawking radiation \cite{unruh1981experimental}. There, the spatiotemporal metric is formed by a background flow in a Bose-Einstein condensate and the analog Hawking radiation originates from a parametric instability \cite{recati2009bogoliubov}. At least in this example, it seems that there is a connection between the Lyapunov exponents and the effective spatiotemporal metric.

\section{Conclusions and Outlook \label{sec:concl}}

I have proposed a quantum trajectory scheme that allows to compute the dynamics of the weakly interacting Bose gas under arbitrary conditions. It consists of the evolution of the mean field and small fluctuations around it. When the fluctuations become large, they are converted into stochastic contributions to the classical field.

This classicization approximation was argued to be valid because the classical field decoheres the quantum fluctuations. The proportionality of the decoherence with the corrections to the Bogoliubov approximation, leads to the fortunate situation that the decoherence approximation becomes good as soon as the corrections to the Bogoliubov approximation start to grow. This makes it possible to keep the Bogoliubov approximation accurate at all times.

The polynomial scaling of the method with system size makes it a promising candidate for the simulation of dynamics in ultracold atomic gases, but the implementation of the model is deferred to future work. Finally, it was argued that the method can be extended to systems whose ground state has a good variational description.

\section*{Acknowledgements}
I thank Lennart Fernandes, Dries Sels, Jacques Tempere, Sonja Hohloch and Wojciech De Roeck for stimulating discussions.

\bibliography{biblio}

\appendix

\section{Normal mode decomposition of a Gaussian state \label{ap:norm}}

The correlation matrix in the quadrature variables is
\begin{equation}
    \sigma^{(xp)}_{ij} =  \langle \{ \delta  \hat r_i, \delta  \hat r_j^\dag \} \rangle =\bar U\sigma^{(a)} \bar U^\dag.
    \label{eq:cortfap}
\end{equation}
It can be written in its symplectic normal form as
\begin{equation}
     \sigma^{(xp)}_{ij} = S D S^T.
     \label{eq:sigmaxpnorm}
\end{equation}
Here, the matrix $S$ is symplectic, i.e. $\Omega$: $S \Omega S^T = \Omega$, with the symplectic form in $x,p$ basis
\begin{equation}
    \Omega = \begin{pmatrix}
        0 & \mathds{1}  \\ -\mathds{1}  & 0
    \end{pmatrix}.
\end{equation}
The symplectic matrix $S$ can in turn be decomposed with a singular value decomposition as
\begin{equation}
    S=O_1 \Sigma O_2^\dag,
    \label{eq:Ssvd}
\end{equation}
where $O_{1}$ and $O_2$ are symplectic and orthonormal. The diagonal matrix $\Sigma$ has the structure
\begin{equation}
\Sigma={\rm diag}(z_1, 1/z_1, z_2,1/z_2,\cdots)    
\end{equation}
contains the squeezing parameters $z_i$ (compare to Eqs. (\ref{eq:sigmax},\ref{eq:sigmap}) in the single mode case).

For the vacuum state, the correlation matrix is equal to the identity, see Eq.~\eqref{eq:sigmablock}. Since any pure state is related to the vacuum by a symplectic transformation, the matrix $D$ in \eqref{eq:sigmaxpnorm} is then the unit matrix. Together with \eqref{eq:Ssvd}, this means that for a pure state, the correlation matrix can be written as
\begin{equation}
    \sigma^{(xp)} = O_1 \Sigma^2 O_1^T.
    \label{eq:sigmaxpdiag_app}
\end{equation}
This decompositions shows that the eigenvalues of the correlation matrix of a pure state come in pairs $z_i^2$ and $1/z_i^2$ belonging to modes with fluctuations that are respectively larger and smaller than the classical shot noise fluctuations of a coherent state.

\end{document}